# Outage Behavior of Slow Fading Channels with Power Control Using Noisy Quantized CSIT

Siavash Ekbatani [*] *IEEE Student Member*, Farzad Etemadi,

and Hamid Jafarkhani, *IEEE Fellow*

**Abstract**

The topic of this study is the outage behavior of multiple-antenna slow fading channels with quantized feedback and partial power control. A fixed-rate communication system is considered. It is known from the literature that with error-free feedback, the outage-optimal quantizer for power control has a circular structure. Moreover, the diversity gain of the system increases polynomially with the cardinality of the power control codebook. Here, a similar system is studied, but when the feedback link is error-prone. We prove that in the high-SNR regime, the optimal quantizer structure with noisy feedback is still circular and the optimal Voronoi regions are contiguous non-zero probability intervals. Furthermore, the optimal power control codebook resembles a channel optimized scalar quantizer (COSQ), i.e., the Voronoi regions merge with erroneous feedback information. Using a COSQ, the outage performance of the system is superior to that of a no-feedback scheme. However, asymptotic analysis shows that the diversity gain of the system is the same as a no-CSIT scheme if there is a non-zero and non-vanishing feedback error probability.

**Index terms**

Outage probability, noisy feedback, channel optimized quantizers, bit-mapping, diversity.

## I. INTRODUCTION

For delay-unconstrained transmission over wireless fading channels, transmission data codewords can span over an infinite number of independent fading realizations. Consequently, ergodic

This work was supported in part by an ARO Multi-University Research Initiative (MURI) grant # W911NF-04-1-0224.

The Authors are with the Center for Pervasive Communications and Computing, Department of Electrical Engineering and Computer Science, University of California, Irvine. Email: { sekbatan , fetemadi , hamidj } @ uci . edu.

This work was presented in part in the IEEE Global Communications Conference, Wasgington DC, November 2007.

capacity is a valid performance measure [1]. In practice, however, delay requirements may limit the transmission to a finite number of fading blocks. Therefore, block fading channel models are more appropriate [2]. In this work, we are interested in very slow fading scenarios where each codeword is transmitted over a single fading block. This channel model is also called quasi-static in the communications literature. In this case, the conventional performance measure is the outage probability [3], [4]. Over the duration of one codeword, the channel coefficients remain constant. Acquiring perfect receiver channel state information (CSI) in this case is relatively simple. Because by definition, we have a long fading block length and accurate channel estimation through training is possible. Obtaining transmitter CSI (CSIT), however, requires feedback in the system. The optimal outage probability of this system with perfect feedback is studied thoroughly in the literature and it is known that power control at the transmitter significantly reduces the outage probability [3], [4].

The more practical scenario where the feedback information is resolution constrained or rate-limited is also studied in the literature using power control approaches [5]–[10]. In this case, to enable power control in the system, a quantizer is defined at the receiver/transmitter ends, so that the receiver can convey an index from a quantization codebook back to the transmitter that represents the state of the channel. Kim *et al.* proposed a quantizer structure for power control in [7], [10] and pointed out that both "strongest" and "weakest" channel realizations must be represented by the same index. In other words, the quantizer structure is circular. Moreover, they proved that with error-free feedback, the optimal quantizer structure has contiguous Voronoi regions.

Quantized feedback is also extensively used in the literature for beamforming over multiple-antenna channels and elegant beamforming strategies such as Grassmannian beamforming have been introduced [11], [12]. These beamforming schemes do not require any power adaptation at the transmitter, however the maximum diversity gain that they can provide is equal to the diversity gain of a no-CSIT system. Note that with the outage probability performance measure, the diversity gain is defined as the decay rate of the outage probability with respect to the signal-to-noise ratio (SNR).

Here, we study a more practical scenario, where not only is the feedback link resolution constrained, but also the feedback indices are subject to errors. This situation naturally arises in a variety of applications since feedback information must also be carried through a fading channel with limited power resources. Moreover, communicating feedback information is severely time-sensitive, and the number of feedback channel uses must be limited. From a practical point of view, we need to minimize the delay between the channel estimation and quantization instance at the receiver and the feedback signal detection instance at the transmitter. Because the transmitter can only adapt the transmission codeword power level, upon receiving (decoding) the feedback index and within the time of communicating feedback, it should remain idle. These limitations of the feedback stage of a closed-loop communication system justifies a noisy channel model for the feedback link.

For simplicity, we consider a discrete memoryless channel (DMC) model of the feedback link, where the bit error (cross-over) probabilities are constant for all channel realizations. A more elaborate finite-state feedback channel model can be found in [13] in a different context. Since we consider slow fading channels, we assume that the channel remains constant during the transmission of $N$ consecutive data symbols. Moreover, we assume that $N$ is large enough so that a rate equal to the instantaneous mutual information of the channel is achievable. We won't consider any feedback delay in our study. We will assume that while the feedback information reaches the transmitter, the channel remains constant, so it does during transmission of one codeword of $N$ channel uses.

Throughout the paper, we also assume that the transmission rate is constant and is known to the receiver. Therefore, regardless of the noisy feedback index appearing at the transmitter, the receiver attempts to decode the received codeword from a codebook of known rate. If the transmission power is greater than or equal to the power needed for successful (outage-free) transmission or in other words, channel inversion, then the decoding succeeds. Otherwise outage occurs. This is in contrast to the assumptions made in rate-adaptive strategies, where the receiver needs to know the feedback index that appears at the transmitter [13], [14].

To design the optimal CSI quantizer for this system, we solve an outage minimization problem under an average (long-term) power constraint at the transmitter. Interestingly, the structure of our quantizer becomes similar to a channel optimized scalar quantizer (COSQ) [15]–[18]. As the quality of the feedback link degrades, the reconstruction points of the power control codebook merge and the outage performance of the system approaches the no-CSIT performance [20].

It has been also shown that the diversity gain of a system with power control based on quantized feedback increases polynomially with the cardinality of the power control codebook if the feedback channel is error-free [5], [7]–[10]. This result motivates us to study the diversity gain of our transmission scheme, where the feedback information is erroneous. Through asymptotic analysis, we show that the achievable diversity gain of the system is equal to the diversity gain of a no-CSIT scheme. Note, however, that by proper design of the quantizer, we can achieve less outage probabilities compared to a no-CSIT scheme.

The rest of the paper is organized as follows. In Section II, we introduce the forward link transmission and the feedback channel model. Numerical design of the quantizer is proposed in Section III and the optimality of the proposed quantizer structure is proven in Section IV. In Section V, we present the asymptotic analysis of the outage probability with noisy feedback and derive the diversity gain of the system. In Section VI, the outage performance of the system is numerically evaluated. Finally, we draw the major conclusions of the work in Section VII.

*Notations:* In the following sections, we use $\mathcal{P}$ for power-related variables and $\mathrm{P}$ or $\mathrm{p}$ to represent probabilities. Furthermore, all the logarithms are natural unless otherwise stated. For two sets $A$ and $B$, $A - B$ represents the set of the members of $A$ that do not intersect with the set $B$.

## II. System Model

### A. Forward Link Transmission Model

The block diagram of the system under study is shown in Fig. 1. In this scheme, the $t$-antenna transmitter scales a rate $\mathrm{R}$ nats per channel use $t$-dimensional complex Gaussian codeword $\mathcal{C}$ by the CSIT-dependent power loading factor $\sqrt{\mathcal{P}_i/t}$. This power control parameter is dictated by the receiver via the feedback index $i$. The covariance matrix of the transmission codeword is $\mathbb{E}\left\{\mathcal{C}\mathcal{C}^{\dagger}\right\} = I_t$, the identity matrix, where $\mathbb{E}\{\cdot\}$ denotes ensemble expectation and $\dagger$ represents the complex conjugate transpose operation.

The elements of the received signal in each channel codeword can be represented by

$$y(l) = H\, x(l) + n(l) \qquad \forall l \in \{1, \cdots, N\} \tag{1}$$

where $l$ denotes the time index within a block of length $N$ channel uses. In this model, $n(l)$ is a unit-power circularly symmetric complex Gaussian noise process and H is the $r \times t$ complex Gaussian channel matrix with entries of variances $0.5$ per real dimension. Furthermore, $r$ is the number of the receive antennas. The channel coefficients remain constant over the duration of one fading block. Assume that the receiver is capable of estimating the channel matrix perfectly and there is a feedback link carrying a limited number of feedback bits from the receiver to the transmitter.

Using a scaled identity covariance codeword with transmission power $\mathcal{P}$, the mutual information between the transmitter and the receiver data can be expressed as

$$I(\mathcal{P}) \triangleq \sum_{\kappa=1}^{\min\{r,t\}} \log\left(1 + \lambda_\kappa \frac{\mathcal{P}}{t}\right) \tag{2}$$

where $\{\lambda_\kappa\}$, $k = 1, 2, \cdots, \min\{r,t\}$ denote the eigenvalues of $HH^{\dagger}$ [7]. For future references, let us define $\mathcal{P}_{\mathrm{R}}(H)$ to be the solution of

$$I(\mathcal{P}) - R = 0 \tag{3}$$

In other words, $\mathcal{P}_{\mathrm{R}}(H)$ is the minimum power level required at the transmitter to invert the channel or the required transmission power so that the receiver can successfully decode the codeword of rate R nats per channel use.

*B. Quantizer Structure and Feedback Channel Model*

The goal of the feedback link is to enable power control in the system. Suppose that the transmitter and the receiver employ a finite-sized power control codebook $\{\mathcal{P}_0, \cdots, \mathcal{P}_{K-1}\}$. The receiver chooses a member of this codebook based on the realization of the channel and conveys its index back to the transmitter using $\lceil \log_2 K \rceil$ feedback bits, where $\lceil \cdot \rceil$ shows the smallest integer greater than or equal to a real number. Inspired by the unlimited feedback results of [3], Kim *et al.* proposed the following feedback index mapping to minimize the probability of outage at rate R, when the feedback link is error-free [8], [10]:

$$\mathcal{J}(\mathrm{H}) = \begin{cases} j & \text{if } \mathcal{P}_{\mathrm{R}}(\mathrm{H}) \in (\mathcal{P}_{j-1}, \mathcal{P}_j] \\ 0 & \text{if } \mathcal{P}_{\mathrm{R}}(\mathrm{H}) > \mathcal{P}_{K-1} \end{cases} \qquad (4)$$

Here, for notational convenience, $\mathcal{P}_{-1} = 0$ and $j \in \{0, \cdots, K-1\}$. Also, the long-term power constraint at the transmitter is expressed as $\mathbb{E}_{\mathrm{H}}\{\mathcal{P}_{\mathcal{J}(\mathrm{H})}\} \leq \mathrm{SNR}$. Since in (4) we allocate $\mathcal{P}_j$ for all $\mathcal{P}_{\mathrm{R}}(\mathrm{H}) \in (\mathcal{P}_{j-1}, \mathcal{P}_j]$, we can view this mapping as a scalar quantizer of $\mathcal{P}_{\mathrm{R}}(\mathrm{H})$, where each reconstruction point coincides with the upper boundary.

In order to utilize a noisy feedback channel, we first propose a heuristic scalar quantizer structure for $\mathcal{P}_{\mathrm{R}}(\mathrm{H})$, and later on, we prove its optimality conditions. In Fig. 2, the quantizer boundaries are shown by $\{\mathcal{Q}_j\}$ and the reconstruction points are represented by $\{\mathcal{P}_j\}$. The set of reconstruction points $\{\mathcal{P}_j\}$ includes the members of a power control codebook that is known to the transmitter/receiver sides. Among $j = \{0, \cdots, K-1\}$, the receiver chooses index $j$ if $\mathcal{P}_{\mathrm{R}}(\mathrm{H}) \in (\mathcal{Q}_{j-1}, \mathcal{Q}_j]$ and sends it to the transmitter through the feedback link. The appropriate index for the region $\mathcal{P}_{\mathrm{R}}(\mathrm{H}) > \mathcal{Q}_{K-1}$ will be discussed in Section III. Note that we won't claim the optimality of this structure in this section.

Each feedback index $j$ at the receiver is first permuted to an index $j'$ using a one-to-one mapping $j' = \delta(j)$, and then each bit of $j'$ is sequentially transmitted over the feedback link. Therefore, the feedback bits are binary representation of index $j'$. We call the mapping block $\delta(j)$ in conjunction with the required binary bit transformation as a bit-mapping scheme. At the transmitter side, a demapping block is used after decoding the feedback bits and converting them into feedback indices.

Note that the indices of the proposed power control codebook are not necessarily sequential in their binary representations. When the feedback is error-prone, the structures of the mapping/demapping blocks should affect the outage probability of the system. As we will discuss in the following sections, the structure of the optimal quantizer depends on the bit-mapping scheme that we use at the feedback link.

The output of the feedback link is an index $i$ and the transmitter chooses its power control factor $\mathcal{P}_i$ upon receiving this index. The indices $i$ and $j$ may be different due to the noise in the feedback link. Assume that the feedback channel is a DMC defined by $\lceil \log_2 K \rceil$ uses of a binary symmetric channel (BSC) with cross-over probability $0 \leq \rho \leq 0.5$, that is known a priori. Note that if $\rho > 0.5$, then the transmitter can simply flip the feedback bits and obtain better results. Therefore, the case of $\rho > 0.5$ is not considered here. Moreover, $\rho = 0.5$ is equivalent to a no-CSIT condition. The conditional index transition probability matrix of the feedback link can be expressed by the elements of its $j'$-th row and $i'$-column as

$$\mathrm{p}(i'|j') = \rho^{d(i',j')}(1-\rho)^{\lceil \log_2 K \rceil - d(i',j')} \tag{5}$$

where $d(i', j')$ denotes the Hamming distance between the binary representations of $i'$ and $j'$.

## III. QUANTIZER OPTIMIZATION WITH THE HEURISTIC STRUCTURE

Our goal in this section is to optimize $(\{\mathcal{Q}_i\}, \{\mathcal{P}_i\})$ for the quantizer structure in Fig. 2 to minimize the outage probability of the system under a long-term power constraint. We won't claim the optimality of this structure in this section and the optimality discussions are postponed to the next section.

First, we start with the feedback index assignment for region $\mathcal{P}_{\mathrm{R}}(\mathrm{H}) > \mathcal{Q}_{K-1}$. In this region, regardless of the index $j$, transmission incurs outage. Therefore, the receiver must choose a feedback index that results in the least transmission power consumption, noting that outage is inevitable in this region. Note that when the feedback is noisy, the probability of receiving a correct index $j$ at the transmitter is greater than the probability of receiving any other index $i \neq j$. Since,

$$(1-\rho)^{\lceil \log_2 K \rceil} \geq (1-\rho)^{\lceil \log_2 K \rceil - n} \rho^n \tag{6}$$

for any $n < \lceil \log_2 K \rceil$ and $\rho \leq 0.5$. The left hand side of the latter inequality is the probability of receiving the correct index and the right hand side is the probability of having $n$-bit errors. Therefore, by assigning the index corresponding to the least power level in the reconstruction codebook, i.e., index $0$ to the region $\mathcal{P}_{\mathrm{R}}(\mathrm{H}) > \mathcal{Q}_{K-1}$, we can minimize the most probable transmission power when feedback is noisy and outage is inevitable. [1] Note that we cannot advise switching-off the transmitter, since this requires an additional index in the reconstruction codebook.

To proceed with the optimization process, let us define the complimentary outage probability at rate R and with the transmission power $\mathcal{P}$ as

$$F(\mathcal{P}) \triangleq \Pr[\,I(\mathcal{P}) \geq R\,] \quad (7)$$

Using this definition, the outage probability of the proposed system can be expressed as

$$P_{\text{out}} = \Pr[\mathcal{P}_R(H) > \mathcal{Q}_{K-1}] + \sum_{i=0}^{K-1} \Pr[i; \mathcal{P}_R(H) \leq \mathcal{Q}_{K-1}; \text{outage}]$$

$$= [1 - F(\mathcal{Q}_{K-1})] + \sum_{i=0}^{K-1} \left\{ p(i|i)[F(\mathcal{Q}_i) - F(\mathcal{P}_i)] + \sum_{j=i+1}^{K-1} p(i|j)\,[F(\mathcal{Q}_j) - F(\mathcal{Q}_{j-1})] \right\} \quad (8)$$

Also, the average transmission power can be expressed as

$$\mathcal{P}_{\text{avg}} = \sum_{i=0}^{K-1} \left\{ \Pr[\mathcal{P}_R(H) > \mathcal{Q}_{K-1}]\,p(i|0) + \sum_{j=0}^{K-1} \Pr[\mathcal{P}_R(H) \in (\mathcal{Q}_{j-1}, \mathcal{Q}_j]] p(i|j) \right\} \mathcal{P}_i$$

$$= \sum_{i=0}^{K-1} \left\{ [1 - F(\mathcal{Q}_{K-1})]p(i|0) + \sum_{j=0}^{K-1} [F(\mathcal{Q}_j) - F(\mathcal{Q}_{j-1})]p(i|j) \right\} \mathcal{P}_i \quad (9)$$

The constrained outage minimization problem can then be stated as follows:

$$\begin{aligned} & \min_{\{\mathcal{P}_i\},\{\mathcal{Q}_i\}} && P_{\text{out}} \\ & \text{s.t.,} && \mathcal{P}_{\text{avg}} - \text{SNR} \leq 0 \\ & \text{and} && \mathcal{P}_i - \mathcal{Q}_i \leq 0\,,\quad \mathcal{Q}_{i-1} - \mathcal{P}_i \leq 0 \end{aligned} \quad (10)$$

In the following lemma, we investigate the properties of the above optimization problem.

*Lemma 1:* For a given set of boundaries, $\{\mathcal{Q}_j\}$ in the heuristic structure, the reconstruction points of the optimal quantizer coincide with the upper boundaries at high-SNR. Namely, as $\text{SNR} \to \infty$, the optimal solutions to the above optimization problem should satisfy: $\mathcal{P}_j^o = \mathcal{Q}_j^o$, $\forall j$.

*Proof:* The proof is provided in Appendix A. □

As discussed in the proof of this lemma, the KKT analysis justifies the structure of our proposed quantizer. Generally speaking, however, we cannot claim the global optimality of the proposed quantizer unless we show the convexity of the optimization problem. The optimality of the quantizer structure will be shown in the next section using a different approach.

---
[1] A more formal proof of the optimality of the quantizer structure will be provided in the next section.

Using the proposed quantizer structure, we proceed with the design of the quantizer. Using Lemma 1, the outage probability of the proposed transmission scheme can be expressed as

$$\mathrm{P}_{\mathrm{out}} = [1 - \mathrm{F}(\mathcal{P}_{K-1})] + \sum_{i=0}^{K-1} \sum_{j=i+1}^{K-1} \mathrm{p}(i|j) \left[\mathrm{F}(\mathcal{P}_j) - \mathrm{F}(\mathcal{P}_{j-1})\right] \tag{11}$$

and the average transmission power $\mathcal{P}_{\mathrm{avg}}$ can be written as

$$\mathcal{P}_{\mathrm{avg}} = \sum_{i=0}^{K-1} \left\{ [1 - \mathrm{F}(\mathcal{P}_{K-1})]\mathrm{p}(i|0) + \sum_{j=0}^{K-1} [\mathrm{F}(\mathcal{P}_j) - \mathrm{F}(\mathcal{P}_{j-1})]\mathrm{p}(i|j) \right\} \mathcal{P}_i \tag{12}$$

Now, the simplified power control codebook design objective can be expressed as

$$\begin{aligned} \min_{\{\mathcal{P}_i\}} \quad & \mathrm{P}_{\mathrm{out}} \\ \mathrm{s.t.} \quad & \mathcal{P}_{\mathrm{avg}} \leq \mathrm{SNR} \end{aligned} \tag{13}$$

In Section VI, we numerically solve the above constrained optimization problem.

In the next section, we investigate the optimality conditions of the proposed quantizer structure and show that this structure is optimal given a certain class of bit-mapping schemes is utilized at the feedback link.

## IV. Optimal Quantizer Structure

The proposed quantizer structure of the previous section has the property that all the Voronoi regions, except for one, are contiguous intervals. For this particular quantizer structure, we optimized the reconstruction points and proved that at high SNR, the reconstruction points also define the boundaries. In this section, we find the optimal partitioning of the space of channel realizations that minimizes the outage probability under an average power constraint. In other words, we remove the contiguity assumption of the Voronoi regions and consider a general form of Voronoi regions as arbitrary unions of sets. The goal is to show that with erroneous feedback, under certain conditions, the proposed quantizer structure in (4) is, in fact, optimal.

## A. Bit-Mapping for the Optimal Quantizer Structure

Generally speaking, the optimal quantizer structure depends on the mapping of quantization indices to binary bits (bit-mapping). In the subsequent sections, we will prove the optimality of the quantizer structure for a certain class of bit-mappings. More specifically, we assume that the bit-mapping in use satisfies the following properties:

$$\forall \ell < j \ : \ \sum_{k=j}^{K-1} \mathrm{p}(k|j) > \sum_{k=j}^{K-1} \mathrm{p}(k|\ell) \tag{14}$$

$$\forall \ell > j \ : \ \sum_{k=j}^{K-1} \mathrm{p}(k|j) \geq \sum_{k=j}^{K-1} \mathrm{p}(k|\ell) \tag{15}$$

$$\forall \ell < j \ , \ \forall m \leq K-1 \ : \ \sum_{k=m}^{K-1} \mathrm{p}(k|j) \geq \sum_{k=m}^{K-1} \mathrm{p}(k|\ell) \tag{16}$$

We call such a mapping scheme a quasi-grey bit-mapping and we conjecture that a quasi-grey bit-mapping exists for an arbitrary number of quantization regions, $K$. In the case of $K = 4$, for instance, a quasi-grey bit-mapping can be realized as

$$\begin{array}{lcccc}
\textit{reconstruction points} \ \longmapsto & \mathcal{P}_0 & \mathcal{P}_1 & \mathcal{P}_2 & \mathcal{P}_3 \\
& \downarrow & \downarrow & \downarrow & \downarrow \\
\textit{quasi-grey bit-mapping} \ \longmapsto & 00 & 11 & 10 & 01
\end{array} \tag{17}$$

The index transition probability matrix of the associated DMC can be written as

$$\mathrm{p}_{i'|j'} = \begin{pmatrix}
(1-\rho)^2 & \rho^2 & \rho(1-\rho) & \rho(1-\rho) \\
\rho^2 & (1-\rho)^2 & \rho(1-\rho) & \rho(1-\rho) \\
\rho(1-\rho) & \rho(1-\rho) & (1-\rho)^2 & \rho^2 \\
\rho(1-\rho) & \rho(1-\rho) & \rho^2 & (1-\rho)^2
\end{pmatrix} \tag{18}$$

It is easy to show that properties (14)-(16) hold for the transition probability matrix (18).

A quasi-grey bit-mapping for an arbitrary $K$ may not be unique and it can be found by an exhaustive search. Table I shows the search outcome for different values of $K$. In this table, such a mapping can be obtained from the $b$-bit binary representation of the elements of the *index vector* introduced in the table, where $b = \lceil \log_2 K \rceil$.

*B. Optimal Quantizer Structure*

Recall that for a specific realization of the fading channel $H$, the minimum outage-free power is denoted by $\mathcal{P}_R(H)$ and is given as the solution of (3). Suppose that the optimal size $K$ power codebook is given as $\{\mathcal{P}_j\} = \{\mathcal{P}_0, \mathcal{P}_1, \cdots, \mathcal{P}_{K-1}\}$, where we assume, without loss of generality

$$0 \leq \mathcal{P}_0 \leq \mathcal{P}_1 \leq \cdots \leq \mathcal{P}_{K-1} < \infty \tag{19}$$

Now denote the optimal partitioning of the channel space by

$$\Psi^o = \{\, \psi_0^o,\, \psi_1^o,\, \cdots,\, \psi_{K-1}^o \,\} \tag{20}$$

In other words, the optimal quantization scheme at the receiver assigns feedback index $j$ to the channel realizations in partition $\psi_j^o$. The following theorem shows that under certain conditions, the proposed quantizer structure of Section II-B is optimal almost everywhere.

For simplicity, we use the term "zero-probability set" to indicate channel realizations that occur with zero probability.

*Theorem 1:* Suppose that a quasi-grey bit-mapping is used for the $K$-level quantizer. In the high SNR regime, $\psi_j^* - \psi_j^o$ has zero probability, where

$$\psi_j^* = \{H\,;\, \mathcal{P}_{j-1} < \mathcal{P}_R(H) \leq \mathcal{P}_j\}\,,\ j \in \{1,\cdots,K-1\} \tag{21}$$

$$\psi_0^* = \{H\,;\, 0 \leq \mathcal{P}_R(H) \leq \mathcal{P}_0\} \cup \{H\,;\, \mathcal{P}_{K-1} < \mathcal{P}_R(H)\} \tag{22}$$

*Proof:* The proof is provided in Appendix C. [2] □

## V. ASYMPTOTIC ANALYSIS

In this section, the goal is to characterize the behavior of the outage probability of the proposed system at very high-SNR. In [7]- [10], Kim *et al.* determined the high-SNR behavior of the system with error-free feedback using the definition of the diversity gain,

$$d = \lim_{\text{SNR} \to \infty} \frac{-\log P_{\text{out}}}{\log \text{SNR}} \tag{23}$$

---

[2]The pre-requisites of this proof are provided in Appendix B-1.

They proved that with error-free feedback and for a fixed-rate system, the diversity gain can be expressed as $d = \sum_{k=0}^{K-1}(d_0)^{k+1}$, where $d_0 = rt$ [7]–[10]. The polynomial diversity gain increase with the number of quantization bins of the feedback information was also shown in [5]. Here, we study the diversity gain of the proposed closed-loop transmission scheme with noisy feedback.

***Theorem* 2:** The diversity gain of the proposed transmission scheme, with power control based on noisy quantized feedback is equal to the diversity gain of a no-CSIT system. That is, when $\rho > 0$, we have
$$d = rt \qquad (24)$$
where $r$ and $t$ are the number of receive antennas and the number of transmit antennas, respectively.

*Proof:* The proof is provided in Appendix D. [3] □

## VI. NUMERICAL RESULTS

In this section, we numerically solve the outage minimization problem (13) based on the optimal quantizer structure. For future references, note that over multiple-input single-output (MISO) and single-input multiple-output (SIMO) (thus SISO) Rayleigh fading channels, the complimentary outage probability $\text{F}(\mathcal{P}_j)$ can be derived in closed-form. For example, it is straightforward to show that over MISO channels with $t$ transmit antennas, $\text{F}(\mathcal{P}_j) = 1 - \frac{\Gamma\left[t, \frac{t[\exp(\text{R})-1]}{\mathcal{P}_j}\right]}{(t-1)!}$, where $\Gamma(a, x) = \int_0^x u^{a-1} e^{-u} du$ is the incomplete gamma function. Similarly, over SIMO channels with $r$ receive antennas, $\text{F}(\mathcal{P}_j) = 1 - \frac{\Gamma\left[r, \frac{\exp(\text{R})-1}{\mathcal{P}_j}\right]}{(r-1)!}$. In the MIMO channel case, however, $\text{F}(\mathcal{P}_j)$ cannot be obtained in closed-form and one way of calculating this probability is through Monte Carlo method.

In the first example, we solve (13) for a SISO link with $K = 2$ and $K = 4$ feedback regions. Fig. 3 shows the outage probability versus the average SNR of this system. We can see that the no-CSIT curve shows diversity gain $d_0 = 1$ that is attributed to the diversity gain of a fixed-rate and no-CSIT system. Moreover, the noiseless feedback system shows diversity gain $d = 2$ for $K = 2$ and $d = 4$ for $k = 4$, which coincides with the polynomial diversity gain $d = \sum_{k=0}^{K-1} d_0^{k+1}$ of a fixed-rate system [7]- [10]. Also note that as we increase the number of quantization regions, we achieve lower outage probabilities.

---
[3]The pre-requisites of this proof are provided in Appendix B-2.

Another observation that we can make from this figure is that the diversity gain of the noisy feedback system is the same as the diversity gain of a no-CSIT scheme. However, the outage probability of the noisy feedback system is less than that of the no-CSIT scheme and this superiority is pronounced more for lower feedback error probabilities. Therefore, although noisy feedback couldn't offer any additional diversity gain compared to a no-CSIT scheme, it improved the system performance, as we optimized the CSI quantizer structure by solving (13).

Fig. 4 shows the outage probability of a 2x2 MIMO channel with $K = 2$ feedback regions. Again, the diversity gain of the noiseless feedback system polynomially increases with $K$ and noisy feedback and no-CSIT systems show the same diversity gain. Let us emphasize that noisy feedback doesn't increase the diversity gain of the system. It nevertheless improves the system performance compared to a no-CSIT scenario. Therefore, feedback information is useful even with errors at the feedback link. From this figure, we can also see that the outage performance of the system converges to that of a no-CSIT scheme, when the feedback link degrades severely or when the cross-over probability of the BSC approaches $0.5$. However, the system is never inferior to a no-CSIT scheme. We attribute this property of the system to the structure of the optimal quantizer that we will investigate in more detail in the next example.

In this example, we study a $2 \times 1$ MISO system with $K = 4$ feedback regions. The optimal quantizer structure can be obtained through solving (13). Fig. 5 shows the structure of the optimal quantizer for the above system. The leftmost set of points denote the optimal quantizer structure with error-free feedback and the rightmost one is the no-CSIT solution. As the error probability of the feedback link increases, the reconstruction points of the optimal power control codebook merge and the optimal quantizer solution converges to the no-CSIT solution. This property of the optimal quantizer resembles channel optimized scalar quantizers (COSQs) [13], [15], [16]. In other words, for more noisy channels, the Voronoi regions of the optimal quantizer will shrink and in some cases, the number of resolvable Voronoi regions will diminish.

Note that COSQ is a joint source-channel coding module. In a COSQ structure, source coding and channel coding should interact in their impact on the feedback information. As the cross-over probability of the feedback link increases, the reconstruction points of the power codebook merge and some of the Voronoi regions shrink or vanish. Therefore, with feedback errors, the system degenerates to a closed-loop system with fewer feedback regions (less source-coding rate), although the number of feedback bits are fixed, thus the channel coding role of the COSQ becomes more prominent. This structure is more general than separate quantization and coding of the feedback bits across the reverse link, and thus it should perform better than separation, when the same feedback channel resources are used.

Next, to show the role of bit-mapping in the optimal quantizer structure, we optimized the system using a (sub-optimal) identity bit-mapping, and we also considered the (optimal) quasi-grey bit-mapping scheme. In both cases, we used the proposed circular quantizer structure with contiguous Voronoi regions. According to Fig. 5, the outcomes of the optimization, i.e., the optimal reconstruction points are different for these two cases. Therefore, we conclude the bit-mapping scheme that we choose at the feedback channel does change the structure of the optimal quantizer. The circular and contiguous quantizer structure is only optimal for a quasi-grey bit-mapping scheme, and for other mappings, some other structure may be optimal.

Fig. 6 shows the outage probability of the above $2\text{x}1$ MISO system with $K = 4$ feedback regions. The same diversity results and outage behavior as the results of the previous examples can be also observed here. Moreover, the amount of performance improvement that (optimal) quasi-grey bit-mapping with circular and contiguous quantizer structure can provide compared to the (sub-optimal) identity mapping can be observed in this figure. Note also that for higher quality feedback links, the gain of optimal bit-mapping is more noticeable.

In this figure, we also compare the performance of the power control scheme that is the focus of this work to the popular Grassmannian beamforming that is a limited-rate feedback scheme using channel eigenvector information at the transmitter [11], [12]. We see that noiseless beamforming outperforms a no-CSIT system with the same diversity gain. However, even with a small feedback error probability, the performance of Grassmannina beamforming degrades severely, and becomes even worse than that of a no-CSIT system. This behavior is not surprising, since Grassmannian beamforming is designed for an error-free feedback system. In the light of these observations, we claim that power control is more effective and more robust compared to Grassmannina beamforming. Since as shown in Fig. 6, even noiseless beamforming cannot outperform the optimal power control scheme with a moderate feedback error probability. Of course, the benefits of power control come at a price, which is the complexity of power adaptation at the transmitter, where beamforming can be performed only with a fixed power. However, if there is any possibility of errors in the feedback link, power control becomes more attractive than beamforming to minimize the outage probability of the system.

## VII. Conclusions

In this work, we formulated and optimized the outage probability of transmission over quasi-static fading channels with temporal power control, when a noisy version of quantized channel state information is available at the transmitter. The presented scheme bypassed the requirement of receiver knowledge of the noisy feedback index at the transmitter.

At high-SNR, using a quasi-grey bit-mapping scheme at the feedback link, the optimal power control codebook with a noisy feedback channel belongs to a circular quantizer structure with contiguous Voronoi regions. Also, the optimal quantizer follows a structure similar to channel optimized scalar quantizers. The outage probability of this system is less than that of a no-CSIT scheme and therefore noisy feedback is useful. However, asymptotic analysis of the outage probability shows that noisy feedback cannot offer any additional diversity gain compared to a no-CSIT scheme, if the error probability of feedback doesn't vanish with the SNR.

## VIII. APPENDIX

### A. Proof of Lemma 1

*Proof:* Our goal is to investigate the Karush-Kuhn-Tucker (KKT) optimality conditions. Problem (10) can be simplified using the Lagrange multipliers $\lambda_p$, $\lambda_i^l$, $\lambda_i^u$, and the Lagrangian

$$\mathcal{L} = \text{P}_{\text{out}} + \lambda_p \left(\mathcal{P}_{\text{avg}} - \text{SNR}\right) + \sum_{i=0}^{K-1} \left[\lambda_i^l(\mathcal{Q}_{i-1} - \mathcal{P}_i) + \lambda_i^u(\mathcal{P}_i - \mathcal{Q}_i)\right] \quad (25)$$

Here, the KKT optimality conditions can be written as [19]

$$\begin{aligned}
-\text{p}(i|i)f(\mathcal{P}_i) + \lambda_p \, \text{p}(i) - \lambda_i^l + \lambda_i^u &= 0 & (\text{i}) \\
\lambda_i^l \geq 0 \quad \text{and} \quad \lambda_i^u &\geq 0 & (\text{ii}) \\
\lambda_i^u(\mathcal{P}_i - \mathcal{Q}_i) = 0 \quad \text{and} \quad \lambda_i^l(\mathcal{Q}_{i-1} - \mathcal{P}_i) &= 0 & (\text{iii})
\end{aligned} \quad (26)$$

for $\forall i$, where $f(\mathcal{P}_i)$ is the derivative of $\text{F}(\mathcal{P})$ at $\mathcal{P}_i$ and $\text{p}(i)$ is the marginal probability of index $i$ at the transmitter. Note that $\text{F}(\mathcal{P})$ is a non-decreasing function of $\mathcal{P}$. Therefore, $f(\mathcal{P}_i)$ is always non-negative. Moreover, as $\text{SNR} \to \infty$, the optimization problem (10) approaches an un-constrained-power optimization problem and thus the Lagrange multiplier associated with the power constraint approaches zero, i.e., $\lim_{\text{SNR} \to \infty} \lambda_p = 0$. Since the first and the second terms in (i) are negative and zero in the limit of high-SNR, respectively and $\lambda_i^l \geq 0$, we must have $\lambda_i^u > 0$. Therefore from relation (iii), we conclude that $\mathcal{P}_i - \mathcal{Q}_i = 0$. In other words, at high-SNR, the structural constraint $\mathcal{P}_i - \mathcal{Q}_i \leq 0$ is active. □

### B. Pre-requisites for the proofs of the Theorems

In this section, we will present a few results that will be used in the proofs of the theorems.

*1) pre-requisites of Theorem 1:*

**Lemma 2:** Suppose that the derivative of the complimentary outage probability function $\text{F}(x)$ in (7) is represented by $f(x)$. In the optimal power codebook, we have

$$\lim_{\text{SNR} \to \infty} f\left(\mathcal{P}_{K-1}\right) \mathcal{P}_{K-1} = 0$$

*Proof:* First we prove a general result for a generic function and then we apply the result to prove the lemma. Suppose that a function $h(x)$ is defined for $\forall x > 0$ and it is bounded above by zero, i.e., $h(x) \leq 0, \ \forall x > 0$. Moreover, suppose that the derivative of $h(x)$, that is $h'(x)$ exists over $\forall x > 0$. We want to show that if the limit of $h'(x)$ also exists, it cannot be a positive number. In other words, $\lim_{x \to \infty} h'(x) \leq 0$.

We use contradiction. Let us suppose the contrary, that is

$$\lim_{x \to \infty} h'(x) = m > 0 \Leftrightarrow \forall \delta > 0, \exists N > 0;\ x > N \Rightarrow |h'(x) - m| < \delta \tag{27}$$

Pick $x_0 > N$ and $m_0 < m$ and find $d_0$, such that

$$h(x_0) = m_0 x_0 + d_0 \tag{28}$$

Now, we must prove that

$$\forall x > x_0,\ h(x) > m_0 x + d_0 \tag{29}$$

Again, suppose the contrary, that is

$$\exists x_1 > x_0;\ h(x_1) \leq m_0 x_1 + d_0 \tag{30}$$

Then, according to (28), (30), and the mean value theorem, we have:

$$\exists x' \in (x_0, x_1);\ h'(x') \leq m_0 \tag{31}$$

This is clearly in contradiction to (27). This leads us to show that (29) is true. If (29) is true, then the function $h(x)$ is greater than a linear function for $x > N$ and cannot be bounded. This contradicts the assumption that we began with. To summarize, we showed that if $h(x)$ is bounded above and the limit of its derivative exists, $\lim_{x \to \infty} h'(x) \leq 0$.

Now, define $h(x) \triangleq F(x)x - x$, where $F(\cdot)$ is the complementary outage probability function, defined in (7). Note that $F(x) \leq 1$, therefore, $h(x) \leq 0$. Now, suppose that the derivative of $h(x)$ exists [4]. From the results of the previous argument, we know the following about the derivative of $h(x)$:

$$\lim_{x \to \infty} f(x)x + F(x) - 1 \leq 0 \tag{32}$$

Moreover, we know that $f(x)x$ is non-negative and $\lim_{x \to \infty} F(x) = 1$. The latter two facts and (32) result in

$$\lim_{x \to \infty} f(x)x = 0 \tag{33}$$

To proceed, we use the result of Lemma 3, in the next sub-section, which states that $\mathcal{P}_{K-1} \to \infty$ as $\text{SNR} \to \infty$. Since $F(\mathcal{P}_{K-1}) \leq 1$, we have $F(\mathcal{P}_{K-1})\mathcal{P}_{K-1} \leq \mathcal{P}_{K-1}$. Therefore, we can define the bounded function

$$h(\mathcal{P}_{K-1}) = F(\mathcal{P}_{K-1})\mathcal{P}_{K-1} - \mathcal{P}_{K-1} \tag{34}$$

Moreover, $f(\mathcal{P}_{K-1})\mathcal{P}_{K-1}$ is always non-negative and $\lim_{\mathcal{P}_{K-1} \to \infty} F(\mathcal{P}_{K-1}) = 1$. We can then apply (33) to show that

$$\lim_{\mathcal{P}_{K-1} \to \infty} f(\mathcal{P}_{K-1})\mathcal{P}_{K-1} = 0 \tag{35}$$

---

[4] Note that for multiple-antenna Rayleigh fading channels, $\lim_{p \to \infty} \frac{d}{dp}(F(p)p - p) = \lim_{p \to \infty} f(p)p + F(p) - 1$ exists.

Therefore, the result of the lemma follows. $\square$

*2) Pre-requisites of Theorem 2:* Without loss of generality, throughout the rest of the proofs, we assume that

$$0 \leq \mathcal{P}_0 \leq \mathcal{P}_1 \leq \cdots \leq \mathcal{P}_{K-1} < \infty \tag{36}$$

***Lemma 3:*** In the optimal power codebook, $\mathcal{P}_0 \leq \text{SNR}$ and $\mathcal{P}_{K-1} \geq \text{SNR}$.

*Proof:* We prove this lemma by contradiction. First, assume that $\mathcal{P}_0 > \text{SNR}$. Since $\mathcal{P}_i \geq \mathcal{P}_0$, $\forall i = \{1, \cdots, K-1\}$, we conclude that $\mathcal{P}_i > \text{SNR}$, $\forall i$, and the power constraint (12) will be violated. Second, assume that $\mathcal{P}_{K-1} < \text{SNR}$. Since $\mathcal{P}_i \leq \mathcal{P}_{K-1}$, $\forall i = \{0, \cdots, K-2\}$, we conclude that $\mathcal{P}_i < \text{SNR}$ $\forall i$. The latter inequality requires the expected power of the system to be less than the SNR. Since the outage probability is a decreasing function of the expected power, with this assumption, the outage probability of the system would be greater than that of a system with the expected power SNR and this violates the optimality of the power control codebook. $\square$

The following corollary is a direct result of Lemma 3.

***Corollary 1:*** In the optimal power control codebook, there should be an intermediate index $\varsigma \in \{0, \cdots, K-2\}$, such that

$$\mathcal{P}_0 \leq \cdots \leq \mathcal{P}_\varsigma \leq \text{SNR} \leq \mathcal{P}_{\varsigma+1} \leq \cdots \leq \mathcal{P}_{K-1} \tag{37}$$

## C. Proof of Theorem 1

*Proof:* Let $\psi_{0l}^* = \{\text{H} \ ; \ 0 \leq \mathcal{P}_R(\text{H}) \leq \mathcal{P}_0\}$ and $\Psi_u = \{\text{H} \ ; \ \mathcal{P}_R(\text{H}) > \mathcal{P}_{K-1}\}$. Note that the random variable being quantized is the scaler quantity $\mathcal{P}_R(\text{H})$, and as a result, $\Psi_u$, $\psi_{0l}^*$, and $\psi_j^*$ ($1 \leq j \leq K-1$) each represent a single contiguous interval on the real line. Furthermore, note that once we prove that $\psi_j^* - \psi_j^o$ has zero probability, this shows the optimality of the proposed quantizer, since both $\{\psi_k^*\}_{k=0}^{K-1}$ and $\{\psi_k^o\}_{k=0}^{K-1}$ span the entire space of channel realizations.

We first claim that all channel realizations in $\Psi_u$ belong to a single Voronoi region and that the optimal index for this region is the index $j = 0$ that minimizes the power consumption. To prove this, we note that the channel realizations in this region will certainly experience outage, regardless of the index $i$ appearing at the transmitter. Suppose that we allocate index $\ell > 0$ to this region. The average power consumption with this allocation, given $\Psi_u$ occurs, can be written as $\mathcal{P}_{avg}^\ell = \sum_{k=0}^{K-1} \text{p}_{k|\ell} \mathcal{P}_k$, whereas upon assigning index 0 to $\Psi_u$, the corresponding parameter can be written as $\mathcal{P}_{avg}^0 = \sum_{k=0}^{K-1} \text{p}_{k|0} \mathcal{P}_k$. According to the third property of the quasi-grey bit mappings, (16), for $m = 0$, and noting that $\{\mathcal{P}_j\}$ is a non-decreasing sequence, we can see that $\mathcal{P}_{avg}^0 \leq \mathcal{P}_{avg}^\ell$.

Therefore, it is more power efficient to assign index $0$ to $\Psi_u$. We therefore conclude that in the optimal structure, the index assigned to this partition is $j = 0$.

For simplicity, in the rest of the proof, we exclude $\Psi_u$ from the possible channel realizations, and we only discuss the complement of $\Psi_u$. Moreover, unless otherwise stated, we only consider the subset $\psi_{0l}^*$ of $\psi_0^*$ when dealing with index $j = 0$.

Recall that the optimal set of channel realizations mapped to a given index $j$ is denoted by $\psi_j^o$ for $j \in \{0, 1, \cdots, K-1\}$. In its most general form, $\psi_j^o$ can be any arbitrary union of non-overlapping intervals, besides some sets of zero probabilities (discrete realizations). We only consider the subsets of $\psi_j^o$ with non-zero probabilities. Our goal is to show that $\psi_j^o$ is equivalent to $\psi_j^*$ in a probabilistic sense and as such, it is comprised of a single interval on the real line.[5]

We proceed the proof of the theorem by contradiction. Suppose that for some $j$, the set $\psi_j^* - \psi_j^o$ has a non-zero probability. This set represents all the channel realizations mapped to index $j$ in the proposed quantizer that don't belong to $\psi_j^o$. Such channel realizations, therefore, belong to some $\psi_k^o$, $k \neq j$, since $\{\psi_k^o\}_{k=0}^{K-1}$ spans the entire space of channel realizations. This, in turn, requires at least one of the two sets $\psi_j^-$ or $\psi_j^+$ to have a non-zero probability, where

$$\psi_j^- = (\psi_j^* - \psi_j^o) \cap \left( \bigcup_{k=0}^{j-1} \psi_k^o \right) \tag{38}$$

$$\psi_j^+ = (\psi_j^* - \psi_j^o) \cap \left( \bigcup_{k=j+1}^{K-1} \psi_k^o \right) \tag{39}$$

To show the contradiction, we need to prove that in the optimal quantizer structure, the above two sets must have zero probabilities.

In the first part of the proof, we use contradiction to show that $\psi_j^-$ has zero probability. Assume the contrary. Then (38) requires $\psi_j^-$ to have a non-zero probability subset of $\bigcup_{k=0}^{j-1} \psi_k^o$ with elements in $\psi_\ell^o$ for some $\ell < j$. We can then arbitrarily choose elements of such a subset to create a subset denoted by $\psi_{j\ell}^-$ with an arbitrarily small non-zero probability. In the following, we show that by changing the optimal index $\ell$ of $\psi_{j\ell}^-$ and making some adjustments in the power codebook, the overall outage probability of the system can be held constant while the power consumption can be reduced. This will then contradict the optimality of the quantizer. This is accomplished in two steps. In the first step, the outage probability is reduced by re-indexing $\psi_{j\ell}^-$ and this results in an increase in the power consumption. In the second step, we revert the outage probability to its original value by making a modification in the power codebook and show that this modification reduces the power consumption by an amount more than the power increase of the first step.

---

[5]With an exception of $\psi_0^o$ that is a union of two non-overlapping sets defined in (22).

*Step 1:* Suppose that we change the index of region $\psi_{j\ell}^-$ from $\ell$ to $j$. This means that all channel realizations in $\psi_{j\ell}^-$ will now be assigned index $j$ at the receiver. Since $\psi_{j\ell}^- \subset \psi_j^*$ and $\mathcal{P}_j$ is a non-decreasing sequence according to (36), then any index $k$ received at the transmitter for $\psi_{j\ell}^-$ will result in outage if $k < j$. Thus, $\sum_{k=j}^{K-1} \mathrm{p}(k|j)$ and $\sum_{k=j}^{K-1} \mathrm{p}(k|\ell)$ denote the non-outage probabilities after and before re-indexing, respectively, when the channel realization falls in $\psi_{j\ell}^-$. The quasi-grey bit-mapping property (14) indicates that re-indexing reduces the outage probability of the system by

$$\Delta_{\mathrm{out}}^1 = \Pr(\psi_{j\ell}^-) \left( \sum_{k=j}^{K-1} \mathrm{p}(k|j) - \sum_{k=j}^{K-1} \mathrm{p}(k|\ell) \right) = \Pr(\psi_{j\ell}^-) \, \mathrm{C}_1 \tag{40}$$

for some $0 < \mathrm{C}_1 < \infty$, where $\Pr(\psi_{j\ell}^-)$ is the probability of the channel realization being in $\psi_{j\ell}^-$.

The above re-indexing changes the expected power consumption of the system for region $\psi_{j\ell}^-$ from $\Pr(\psi_{j\ell}^-) \sum_{k=0}^{K-1} \mathrm{p}(k|\ell)\mathcal{P}_k$ to $\Pr(\psi_{j\ell}^-) \sum_{k=0}^{K-1} \mathrm{p}(k|j)\mathcal{P}_k$. The latter difference is given by

$$\Delta_{\mathcal{P}_{avg}}^1 = \Pr(\psi_{j\ell}^-) \sum_{k=0}^{K-1} \left( \mathrm{p}(k|j) - \mathrm{p}(k|\ell) \right) \mathcal{P}_k \tag{41}$$

that is a positive number according to (16) and the non-decreasing property of the power codebook, (36). By using the absolute value of the terms $\mathrm{p}(k|j) - \mathrm{p}(k|\ell)$ and the largest power level, $\mathcal{P}_{K-1}$, in the above equation, we can upper bound $\Delta_{\mathcal{P}_{avg}}^1$ as

$$\Delta_{\mathcal{P}_{avg}}^1 \leq \Pr(\psi_{j\ell}^-) \, \mathrm{C}_2 \, \mathcal{P}_{K-1} \tag{42}$$

for some $0 < \mathrm{C}_2 < \infty$. From (40), we can then conclude that

$$\Delta_{\mathcal{P}_{avg}}^1 \leq \mathrm{C}_3 \, \mathcal{P}_{K-1} \, \Delta_{\mathrm{out}}^1 \tag{43}$$

for some $0 < \mathrm{C}_3 < \infty$. Thus, the net effect of re-indexing was to reduce the outage probability and to increase the expected power by the amounts that are related through (43).

*Step 2:* In this step, we show that by changing the quantizer structure, it is possible to 1) revert the outage probability to its original value before the re-indexing of step 1 and 2) save the expected power by an amount more than the increase in the power consumption of step 1, (43). This will then contradict the optimality of the quantizer. For this purpose, we slightly reduce $\mathcal{P}_{K-1}$ to $\mathcal{P}'_{K-1} = \mathcal{P}_{K-1} - \epsilon$, for some $\epsilon > 0$, and investigate the resulting effects on the outage versus expected power performance of the system.

We assume that $\epsilon$ is sufficiently small so that all the channel realizations in $\mathcal{P}_R(\mathrm{H}) \in (\mathcal{P}'_{K-1}, \mathcal{P}_{K-1}]$ are assigned to a single index $m \in \{0, 1, \cdots, K-1\}$. We denote the marked region by $\psi_{K-1,m}$. This power reduction increases the outage probability of the system by

$$\Delta_{\text{out}}^2 = \Pr(\psi_{K-1,m}) \, \mathrm{p}(K-1|m) = \left[\mathrm{F}(\mathcal{P}_{K-1}) - \mathrm{F}(\mathcal{P}'_{K-1})\right] \mathrm{p}(K-1|m) \tag{44}$$

while reducing the power consumption by

$$\Delta_{\mathcal{P}_{avg}}^2 = \Pr[K-1] \left(\mathcal{P}_{K-1} - \mathcal{P}'_{K-1}\right) \tag{45}$$

where $\Pr[K-1]$ is the probability that index $K-1$ appears at the transmitter. It is important to note that in a noisy feedback system, $\Pr[K-1] \neq 0$.

By equating the changes in the outage probability of the two steps, i.e., $\Delta_{\text{out}}^1 = \Delta_{\text{out}}^2$, and from (43), we arrive at the following upper bound for the power increase of step 1:

$$\Delta_{\mathcal{P}_{avg}}^1 \leq \mathrm{C}_3 \left[\mathrm{F}(\mathcal{P}_{K-1}) - \mathrm{F}(\mathcal{P}'_{K-1})\right] \mathrm{p}(K-1|m) \, \mathcal{P}_{K-1} \tag{46}$$

We now show that in the high SNR regime, $\Delta_{\mathcal{P}_{avg}}^2$ is greater than the above upper bound for $\Delta_{\mathcal{P}_{avg}}^1$. To prove this, (45) and (46) require that

$$\Pr[K-1] \left(\mathcal{P}_{K-1} - \mathcal{P}'_{K-1}\right) > \mathrm{C}_3 \left[\mathrm{F}(\mathcal{P}_{K-1}) - \mathrm{F}(\mathcal{P}'_{K-1})\right] \mathrm{p}(K-1|m) \, \mathcal{P}_{K-1} \tag{47}$$

or equivalently,

$$\frac{1}{\mathrm{C}_3} \frac{\Pr[K-1]}{\mathrm{p}(K-1|m)} > \frac{\left[\mathrm{F}(\mathcal{P}_{K-1}) - \mathrm{F}(\mathcal{P}'_{K-1})\right]}{\left(\mathcal{P}_{K-1} - \mathcal{P}'_{K-1}\right)} \mathcal{P}_{K-1} \tag{48}$$

The left hand side of (48) is a bounded, positive real number denoted by $\mathrm{C}_4 \triangleq \frac{1}{\mathrm{C}_3} \frac{\Pr[K-1]}{\mathrm{p}(K-1|m)}$. Since both $\Pr[K-1]$ and $\text{p}(K-1|m)$ are non-zero real numbers. For an arbitrarily small $\epsilon$, the right hand side of (48) converges to $\text{f}(\mathcal{P}_{K-1})\mathcal{P}_{K-1}$, where $\text{f}(x)$ is the derivative of $\text{F}(x)$, assuming that the derivative exists. To prove the inequality, thus, we need to show that

$$\text{C}_4 > \text{f}(\mathcal{P}_{K-1})\mathcal{P}_{K-1} \tag{49}$$

for any given $\text{C}_4$. Corollary 1 shows that $\mathcal{P}_{K-1} \to \infty$ in the high SNR regime, and Lemma 2 consequently proves (49), since the limit of the right hand side of (49) is zero. We can now conclude from (45) and (46) that

$$\Delta^2_{\mathcal{P}_{avg}} > \Delta^1_{\mathcal{P}_{avg}} \tag{50}$$

In other words, the two steps have kept the outage probability intact, while reducing the expected power of the system. This contradicts the optimality of the partitioning $\{\psi^0_j\}$ and the power codebook $\{\mathcal{P}^o_j\}$.

In the second part of the proof, we show that in the optimal quantizer structure, $\psi^+_j$, defined in (39) also has zero probability. We use contradiction once again. Suppose that for some $\ell > j$, the set $\psi^+_j \cap \psi^o_\ell$ has a non-zero probability. We change the index $\ell$ of a subset of this set, denoted by $\psi^+_{j\ell}$, to $j$. According to (15), $\sum_{K=j}^{K-1} \text{p}(k|j) \geq \sum_{K=j}^{K-1} \text{p}(k|\ell)$, and as a result, re-indexing of $\psi^+_{j\ell}$ reduces the outage probability of the system. Furthermore, according to (16) and the non-decreasing property of the power control codebook, we know that

$$\sum_{k=0}^{K-1} \text{p}(k|\ell)\mathcal{P}_k \geq \sum_{k=0}^{K-1} \text{p}(k|j)\mathcal{P}_k \tag{51}$$

Therefore, re-indexing also reduces the average power consumption. Therefore, index $\ell$ cannot be the optimal index assigned to the region $\psi^+_{j\ell}$ and the contradiction follows. In conclusion, in the optimal quantizer structure, we cannot have a non-zero probability set $\psi^+_{j\ell} \subset \psi^+_j \cap \psi^o_\ell$ for any $\ell > j$.

To conclude the proof, note that the above two steps show that $\Psi^o_j \subset \Psi^*_j$ holds with possible exceptions on some zero-probability subsets of $\Psi^*_j$. Moreover, since both $\{\Psi^*_j\}$ and $\{\Psi^o_j\}$ span the entire space of channel realizations, we conclude that $\Psi^*_j \subset \Psi^o_j$. Therefore, $\Psi^o_j = \Psi^*_j$ holds with possible exceptions on some subsets with zero probabilities. □

## D. Proof of Theorem 2

Let us adopt the notation

$$f(\text{SNR}) \doteq \text{SNR}^b \Leftrightarrow \lim_{\text{SNR}\to\infty} \frac{\log f(\text{SNR})}{\log \text{SNR}} = b \tag{52}$$

as the definition of the order of a function of SNR. Operators $\dot{\leq}$ and $\dot{\geq}$ can also be defined similarly.

*Proof:* First, we show that the diversity gain is upper bounded as $d \leq rt$, using the result of the Corollary 1. Suppose that the intermediate index introduced in Corollary 1 is $\varsigma = 0$. In this case, $\lim_{\text{SNR}\to\infty} F(\mathcal{P}_\kappa) = 1$, $\kappa \in \{1, \cdots, K-1\}$ and from (11), the limit of the system outage probability can be expressed as

$$\lim_{\text{SNR}\to\infty} P_{\text{out}} = \lim_{\text{SNR}\to\infty} p(0|1)[1 - F(\mathcal{P}_0)] \tag{53}$$

Now, suppose that $\varsigma = 1$. Then, $\lim_{\text{SNR}\to\infty} F(\mathcal{P}_\kappa) = 1$, $\kappa \in \{2, \cdots, K-1\}$ and the limit of the system outage probability becomes

$$\lim_{\text{SNR}\to\infty} P_{\text{out}} = \lim_{\text{SNR}\to\infty} p(0|1)[F(\mathcal{P}_1) - F(\mathcal{P}_0)] + p(0|2)[1 - F(\mathcal{P}_1)] + p(1|2)[1 - F(\mathcal{P}_1)] \tag{54}$$

$$\geq \lim_{\text{SNR}\to\infty} p'_{min} \left( F(\mathcal{P}_1) - F(\mathcal{P}_0) + 1 - F(\mathcal{P}_1) + 1 - F(\mathcal{P}_1) \right)$$

$$= \lim_{\text{SNR}\to\infty} p'_{min} \left( [1 - F(\mathcal{P}_0)] + [1 - F(\mathcal{P}_1)] \right)$$

$$\geq \lim_{\text{SNR}\to\infty} p'_{min}[1 - F(\mathcal{P}_0)] \tag{55}$$

where $p'_{min} = \min\{p(0|1), p(0|2), p(1|2)\}$. Note that if $\rho > 0$, the term $p'_{min}$ is a positive real number.

Repeating the above arguments for $2 < \varsigma \leq K - 2$, we observe that the high-SNR expression of the outage probability is always greater than or equal to $p'[1 - F(\mathcal{P}_0)]$, where $p'$ is a positive and constant number. Therefore, if $\rho$ doesn't vanish in the limit of high-SNR, we have

$$P_{\text{out}} \dot{\geq} 1 - F(\mathcal{P}_0) \tag{56}$$

Note that $1 - F(\mathcal{P}_0)$ is equivalent to the outage probability of a no-CSIT system with average power $\mathcal{P}_0 \leq \text{SNR}$. Therefore, from [20], we have

$$1 - \text{F}(\mathcal{P}_0) \dotgeq \text{SNR}^{-rt} \qquad (57)$$

From (57), we conclude that the diversity gain of the noisy feedback system is bounded by

$$d \leq rt \qquad (58)$$

Now we prove that the diversity gain $d = rt$ is achievable using an example. Suppose that in a hypothetical power control codebook, $\mathcal{P}_i = \text{SNR}, \forall i = \{0, \cdots, K-1\}$. The latter codebook doesn't violate the power constraint in (12) and is a valid (but not necessarily optimal) solution. Moreover, since with this codebook $\lim_{\text{SNR} \to \infty}[\text{F}(\mathcal{P}_i) - \text{F}(\mathcal{P}_j)] = 0, \ \forall i, j$, from (11), the high-SNR outage probability of this scheme can be expressed as

$$\lim_{\text{SNR} \to \infty} \text{P}_{\text{out}} = \lim_{\text{SNR} \to \infty} [1 - \text{F}(\mathcal{P}_{K-1})]$$

As a result, we can derive the diversity gain of the system with the latter codebook as

$$\lim_{\text{SNR} \to \infty} \frac{-\log \text{P}_{\text{out}}}{\log \text{SNR}} = \lim_{\text{SNR} \to \infty} \frac{-\log[1 - \text{F}(\text{SNR})]}{\log \text{SNR}} = rt \qquad (59)$$

noting that the two limits in the above expression exist. Therefore, the diversity gain $d = rt$ can be achieved using the above codebook. The outage probability of the system employing the optimal power codebook must be less than or equal to the achievable outage probability with the above codebook. Therefore, we have a lower bound on the diversity gain of the optimal system,

$$d \geq rt \qquad (60)$$

From (58) and (60) we conclude that $d = rt$. □

## Acknowledgement

The authors would like to thank Tung T. Kim for his valuable suggestions that improved the quality of the presentation of this work.

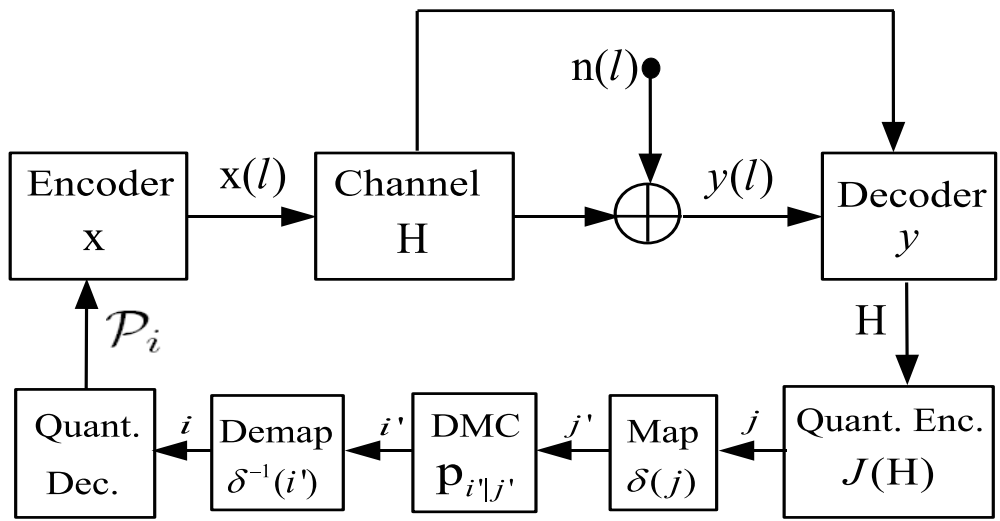

Fig. 1. System block diagram.

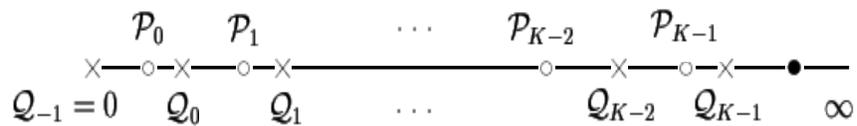

Fig. 2. A heuristic quantizer structure.

TABLE I
QUASI-GREY BIT-MAPPING FOR DIFFERENT QUANTIZER DIMENSIONS.

| dimension | index vector | bits |
|---|---|---|
| $K = 2$ | [ 0  1 ] | $b = 1$ |
| $K = 3$ | [ 0  2  1 ] | $b = 2$ |
| $K = 4$ | [ 0  3  2  1 ] | $b = 2$ |
| $K = 6$ | [ 0  3  5  1  2  4 ] | $b = 3$ |
| $K = 8$ | [ 0  3  6  5  2  7  1  4 ] | $b = 3$ |
| $K = 10$ | [ 0  6  9  7  3  5  1  3  8  2 ] | $b = 4$ |
| $K = 12$ | [ 0  7  2  8  4  9  5  11  1  3  6  10 ] | $b = 4$ |

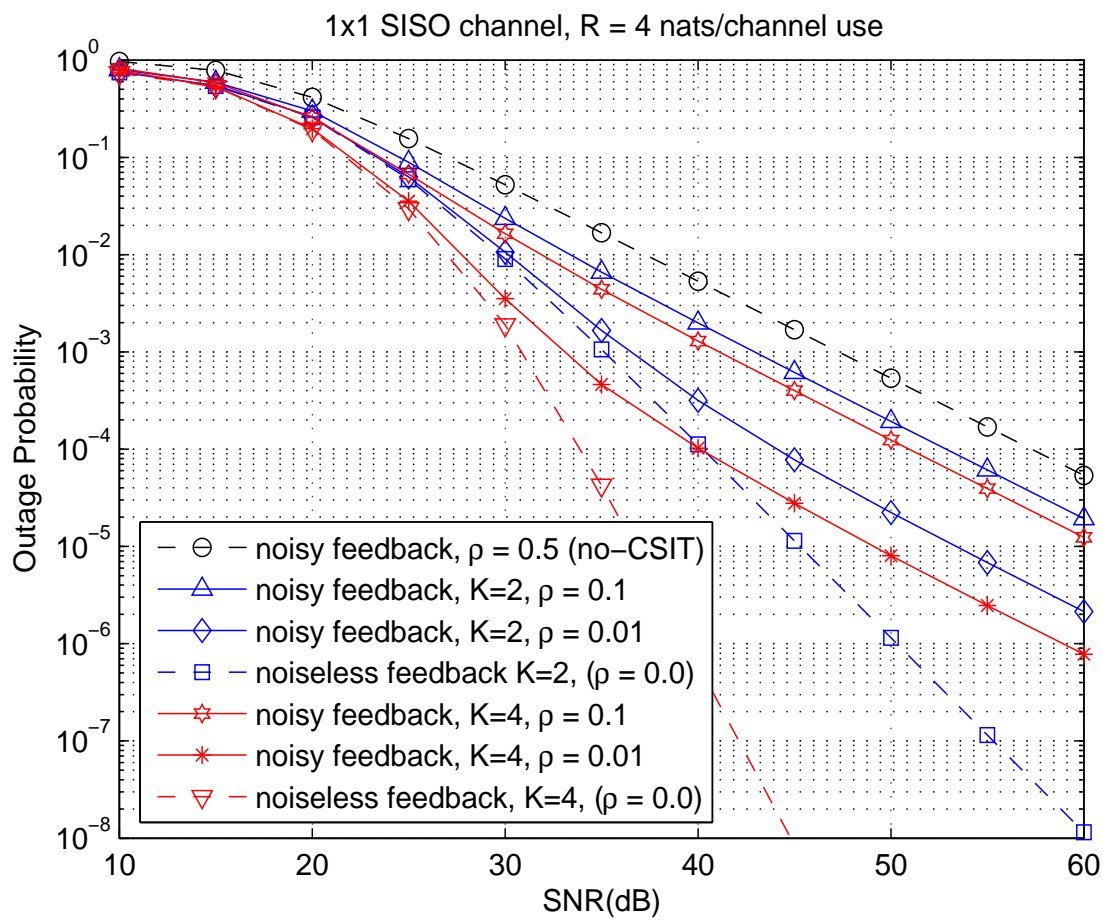

Fig. 3. P_out vs SNR for a $1 \times 1$ SISO channel, with $K = 2$ and $K = 4$ regions, and the transmission rate $R = 4$ nats/channel use.

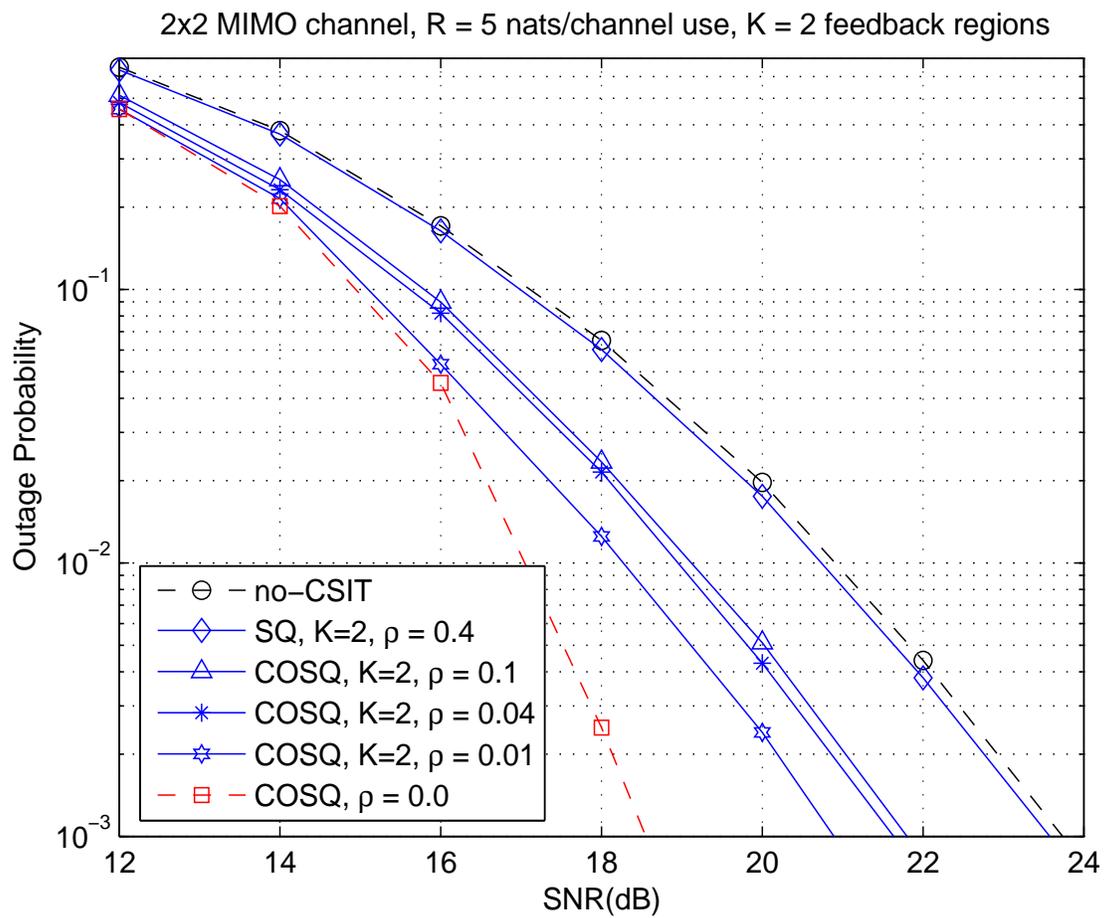

Fig. 4. $P_{\text{out}}$ vs SNR for a $2 \times 2$ MIMO channel, with $K = 2$ regions, and the transmission rate $R = 5$ nats/channel use.

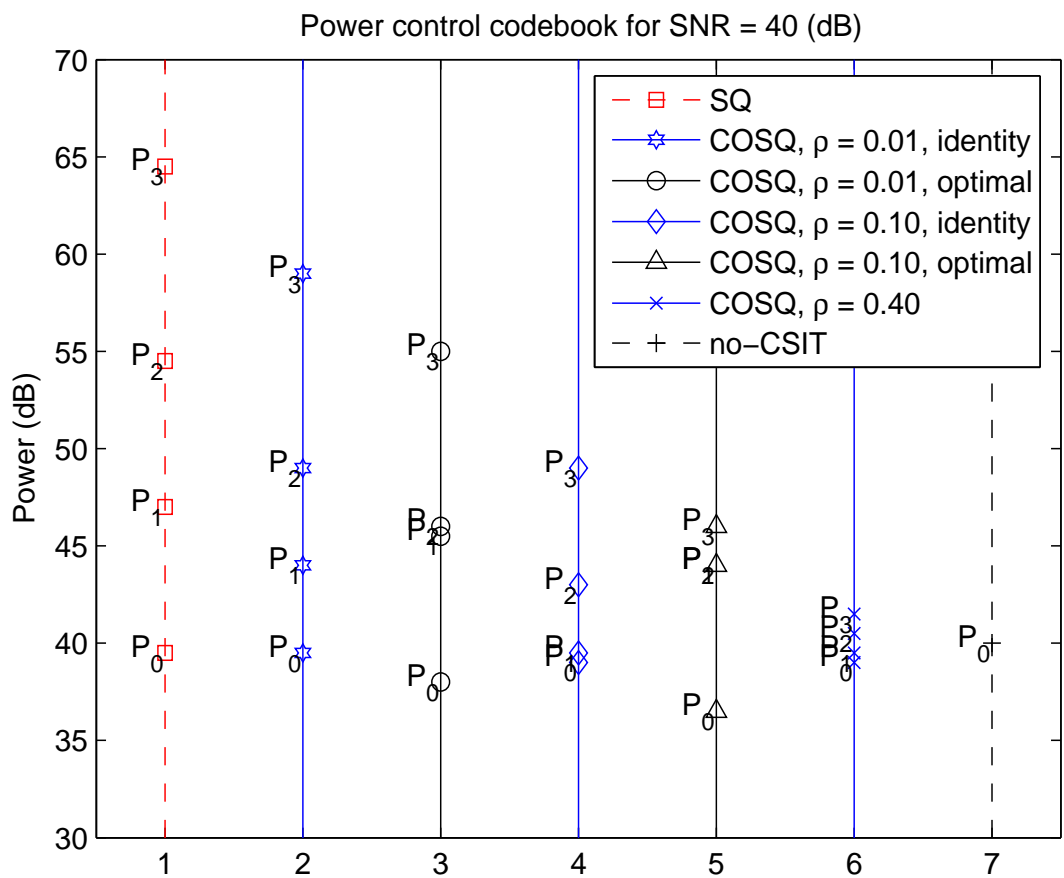

Fig. 5. Optimal power control codebook extracted from numerical optimization. 2x1 MISO channel with $K = 4$ feedback regions.

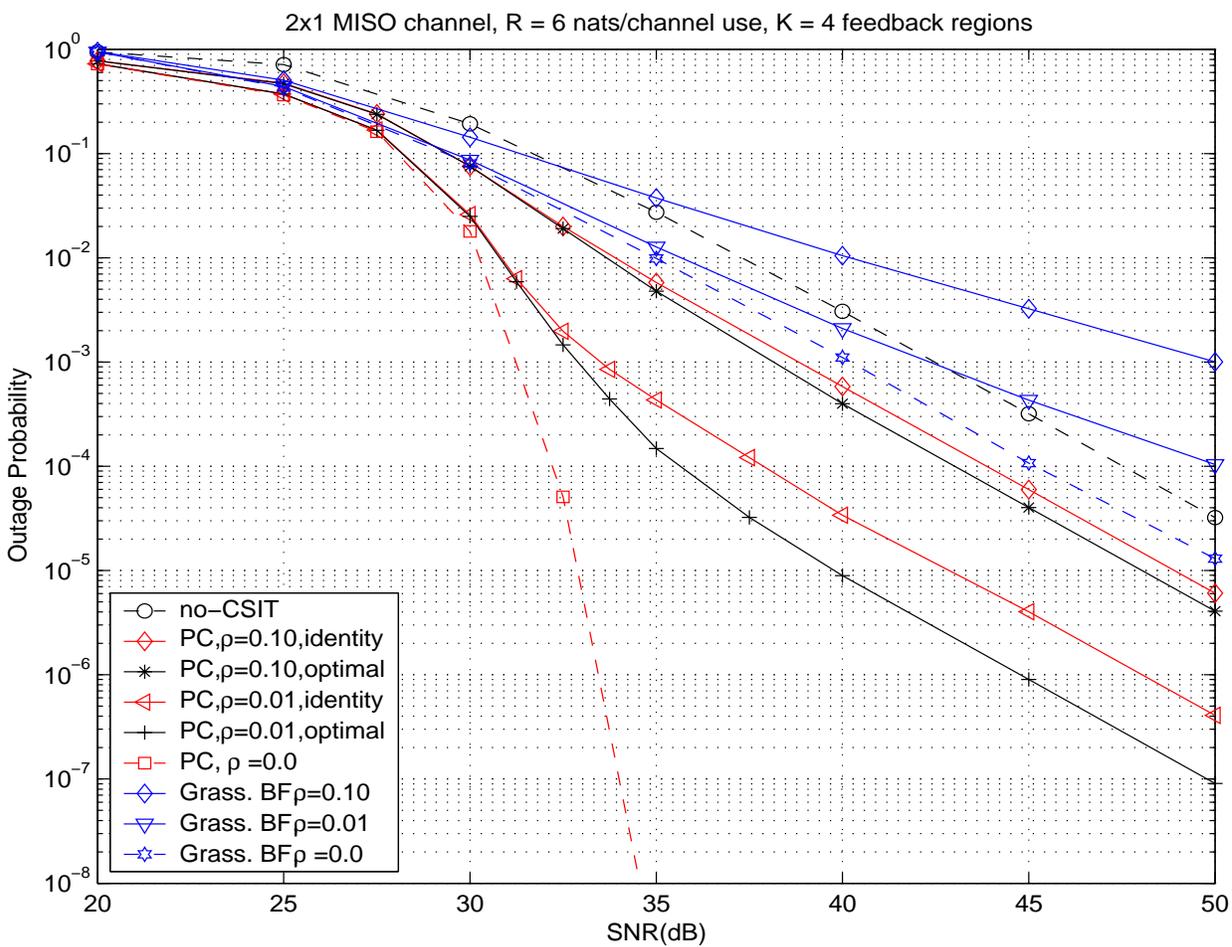

Fig. 6. $P_{\text{out}}$ vs SNR for a $2 \times 1$ MISO channel, with $K = 4$ regions, and the transmission rate R = 6 nats/channel use. Identity index mapping is compared to the optimal index assignment.